\magnification=1200
\baselineskip=24pt

\font\bigbf=cmbx10  scaled\magstep2 \vskip 0.2in
\centerline{\bigbf Noncommutative geometry and } \vskip 0.1in
\centerline{\bigbf the classical orbits of particles in a central
force potential }

\vskip 0.4in \font\bigtenrm=cmr10 scaled\magstep1
\centerline{\bigtenrm  B. Mirza$^{\dag \ddag}$ and M.
Dehghani$^{\dag}$} \vskip 0.2in

\centerline{\sl $^{\dag}$Department of Physics, Isfahan University of Technology, Isfahan 84154, Iran }
\vskip 0.1in
\centerline{\sl $ ^{\ddag}$Institute for Studies in Theoretical Physics and Mathematics, }
\centerline{\sl P.O.Box 5746, Tehran, 19395, Iran}
\vskip 0.1in

\centerline{\sl E-mail: b.mirza@cc.iut.ac.ir}

\vskip 0.2in \centerline{\bf ABSTRACT} \vskip 0.1in

We investigate the effect of the noncommutative geometry on the
classical orbits of particles in a central force potential. The
relation is implemented through the modified commutation
relations $[x_i , x_j]=i \theta_{ij} $. Comparison with
observation places severe constraints on the value of the
noncommutativity parameter.

 \vskip 0.2in
  \noindent PACS numbers: 02.40.Gh, 03.65.Sq, 91.10.Sp, 96.30.Dz

\noindent Keywords: Noncommutative geometry; Planetary orbits.

\vfill\eject

\vskip 1in


Recently, remotivated by string theory arguments, noncommutative
spaces have been studied extensively (for a review see [1,2]). One
might postulate noncommutativity for a number of reasons, perhaps
the simplest is the long-held belief that in quantum theories
including gravity, space-time must change its nature at short
distances. Quantum gravity has an uncertainty principle which
prevents one from measuring positions to better accuracies than
the Planck length: the momentum and energy required to make such a
measurement will itself modify the geometry at these scales [3].
While motivation for this kind of space with noncommuting
coordinates is mainly theoretical, it is possible to look
experimentally for departures from the usually assumed
commutativity among the space coordinates, e.g. see [4,5]. In this
paper we consider the effects of the deformation of the canonical
commutation relations on the orbits of classical particles in a
central force potential. Usual quantum mechanics is formulated on
commutative spaces satisfying the following commutation relations,
$$ [\hat x_i,\hat x_j]=0 \, ,\, \,\,  [ \hat p_i , \hat p_j]=0
\, ,\,\,\,[\hat x_i  , \hat p_j]=i\hbar \delta_{ij} \eqno(1)$$
\noindent
Then in order to describe a noncommutative space, the
above commuatation relations should be changed as,

$$ [\hat x_i,\hat x_j]=i\theta_{ij}  \, ,\, \,\,  [ \hat p_i , \hat p_j]=0
 ,\,\,\,[\hat x_i  , \hat p_j]=i\hbar \delta_{ij} \eqno(2)$$

\noindent where $\theta_{ij} $'s are c-numbers with the
dimensionality $($ length $)^{2}$. In the classical limit, the
quantum mechanical commutator is replaced by the Poisson bracket
via

$${1\over i\hbar}[\hat A ,\hat B] \longrightarrow \{\hat A,\hat B \}\eqno(3) $$
So the classical limit of Eqs. (3) reads

$$ \{\tilde x_i,\tilde x_j\}=\alpha_{ij}  \, ,\, \,\,  \{\tilde p_i , \tilde p_j\}=0
 ,\,\,\,\{\tilde x_i  ,\tilde p_j\}=\delta_{ij} \eqno(4)$$

\noindent  We are keeping the parameters
$\alpha_{ij}={\theta_{ij}\over \hbar }$ fixed as $\hbar
\rightarrow 0 $, for similar arguments see [6].
 The Poisson bracket must possess the same properties as the quantum mechanical
 commutator, namely, it must be bilinear, anti-symmetric and must
 satisfy the Leibniz rules and the Jacobi Identity. The general
 form of the poisson brackets for this deform version of classical
 mechanics can be written as [6,7],

$$\{A,B\}=({\partial A \over  \partial \tilde x_i } {\partial B \over \partial \tilde p_j }-
{\partial A \over \partial \tilde p_i } {\partial B \over \partial
\tilde x_j})\{\tilde x_i,\tilde p_j\}+ {\partial A \over \partial
\tilde x_i } {\partial B \over \partial \tilde x_j}\{\tilde
x_i,\tilde x_j\}\eqno(5) $$ \noindent where repeated indices are
summed. The Hamiltonian of a particle in a central force
potential is given by,

$$ H={\tilde P^2\over 2m} + V(\tilde r) \, \ \ \ \ \  r=\sqrt{\tilde
x_i \tilde x_i} \eqno(6) $$

\noindent To make this situation tractable, one may choose a new
coordinate system,
$$ x_i=\tilde x_i +{1\over 2 }\ \alpha_{ij}\tilde p_j \ , \ \ \
p_i=\tilde p_i \eqno(7) $$

 \noindent
where the new variables satisfy the usual canonical brackets.

$$ \{ x_i, x_j\}=0  \, ,\, \,\,  \{ p_i ,  p_j\}=0
 ,\,\,\,\{ x_i  , p_j\}=\delta_{ij} \eqno(8) $$

\noindent By replacing the new variables in the potential, one has

$$\eqalignno{ V(\tilde r)&= V( \sqrt
{(x_i-\alpha_{ij}p_j/2)(x_i-\alpha_{ik}p_k/2)} \ ) \cr
 &= V(r) + {(\vec \alpha \times \vec p\ )\over 2}\cdot {\vec \nabla V(r)}+ O(\alpha^2 ) \cr
  &= V(r)- {{(\vec \alpha \cdot \vec L)}\over 2 r} {\partial V \over \partial r}+ O(\alpha^2 ) &(9) \cr }  $$

\noindent where $\alpha_{ij}=\epsilon_{ijk} \alpha_k , \  \vec
L=\vec r \times \vec p $. For the Coulomb potential the
Hamiltonian up to the  first order in $\alpha $ becomes,

$$ H= {p^2 \over 2 m } - {k \over r} -{k\over r^3}({{{\vec \alpha \cdot \vec L}\over 2 }}) \eqno(10)$$

\noindent The new term in the Hamiltonian is small and its
effects can be obtained by standard perturbation theory, however
it causes a time dependent angular momentum. We assume that the
time variation of the vector $\vec L $ is so small that in a short
time interval (for example one century) it could be taken as a
constant of motion. Our aim is to put a bound on the value of
$\alpha $ by comparing the results of the perturbing term with the
experimental value of the precession of the perihelion of
Mercury. For the Kepler problem it is known that the bounded
orbits are closed, which is a result of the following integral,

$$ \eqalignno{ \triangle \varphi^{(0)} & = -2 {\partial \over
\partial L}\int_{r_{min}}^{r_{max}} { \sqrt {2 m (E- V(r)) - {L^2 \over r^2
}}}\ \  dr &(11) \cr &= 2 \pi &(12) \cr}$$

\noindent where

$$ E= {1\over 2} m(\dot r^2 + r^2 \dot {\varphi}^2) + V(r) \ , \ \ \ L=m r^2 \dot {\varphi} \eqno(13)$$

\noindent By perturbing the potential with a small term $ V
\rightarrow V + \delta V $ and expanding the integral up to the
first order in $ \delta V $, one has

$$  \triangle \varphi = \triangle \varphi^{(0)} + \triangle
\varphi^{(1)} \eqno(14) $$

\noindent where

$$ \triangle
\varphi^{(1)}= {\partial \over \partial L}{\int_0^\pi ({{2 m
r^2}\over L})\  \delta V \  d\varphi }\eqno(15) $$

\noindent After a straightforward calculation, we arrive at

$$ \triangle
\varphi^{(1)}=( { {2  \pi k^2  m^2   cos( \gamma ) }\over
L^3})\alpha \eqno(16)
$$

\noindent where $\gamma $ is an angle between $\vec \alpha $ and
$\vec L $. According to Ref. [8], the observed advance of the
perhelion of Mercury that is unexplained by Newtonian planetry
perturbations or solar oblateness is

$$ \eqalignno{\triangle \varphi_{obs} &=42.980 \pm 0.002 \ \
\rm{arc \ seconds \  per \  century } \cr &= 2 \pi (7.98734 \pm
0.00037) \times 10^{-8}  \ \rm{ radians/revolution } &(17)\cr } $$

\noindent This advance is usually explained by General Relativity
which predicts,

$$ \triangle \varphi_{GR} =  {{6 \pi G M}\over { c^2 a(1-e^2)}} \eqno(18)$$

\noindent For Mercury, the parameters are [9,10],

$$ {{2 G M} \over c^2} = 2.95325008 \times 10^3 \  m  $$
$$ m=3.3022 \times 10^{23} \ kg $$
$$ e=0.20563069 $$
 $$ \eqalignno{a &= {{r_{max}+r_{min}}\over 2 }\cr &= 5.7909175 \times
10^{10} \ m &(19) \cr} $$

\noindent and

$$ \triangle \varphi_{GR} =2 \pi (7.98744\times 10^{-8} ) \ {\rm
radians/revolution }  \eqno(20)$$

\noindent Comparison of Eqs. (17) and (20) yields

$$  \triangle \varphi_{GR}- \triangle \varphi_{obs}=2 \pi
(0.00010 \pm 0.00037 ) \times 10^{-8} \ \rm{ radians/revolution}
\eqno(21) $$

\noindent  So we can assume that in this scenario,

$${ \left | \triangle  \varphi^{(1)} \right |} \leq {\left |\triangle \varphi_{GR}- \triangle
\varphi_{obs}\right |} \eqno(22) $$

\noindent Thus,

 $$ \alpha \ cos(\gamma ) \leq 1.2 \times 10^{-29} \ \ \ ({m^2\over J. \ s.} ) \eqno(23) $$

\noindent  Considering a scenario in which the angle $\gamma $
takes values from $0 $ to a few seconds less than $ \pi\over 2 $.
So we can place a constraint on the value of the noncommutative
parameter $\theta $

$$  \hbar \  \alpha  = \theta  \leq  10^{-62} \ \  m^2 \eqno(24)$$

\noindent and

$$ \sqrt { \theta } \approx  10^{3} \ell_{planck} \eqno(25)$$

\noindent This limit is much smaller than the limits which have
already been obtained [4,5].  For $ \gamma $ equal to $\pi\over 2
$ the perturbing term vanishes and one has to consider higher
orders in $\alpha $. By a similar calculation with $ \gamma={
\pi\over 2}$ and up to the second order in $ \alpha $ we obtain
an even smaller number for $\alpha $ in this special scenario.

\noindent As we have already mentioned, the perturbing term in Eq.
(10) causes a time dependent angular momentum.

$$ \eqalignno{ {d \vec L  \over dt}&=\{ \vec L , H \} \cr
&= -{{k}\over  r^3}\ ( {{\vec \alpha \times \vec L }\over 2} )
&(26)\cr }$$

\noindent Eq. (26) has a simple physical interpretation, the
angular momentum vector  $\vec L $ rotates around $\vec \alpha $
with a frequency which is about,

$$ w_{rot}\simeq ({G M m   \over 2  r^3 \hbar }) {\theta } \eqno(27) $$

\noindent It means that in this scenario, the two dimensional
plane which contains Mercury and the sun and is perpendicular to
the angular momentum vector has a rotation about $\vec \alpha $
with a period of about ten billion $ (10^{10})  $ years which is
indeed a long time and comparable to the age of the solar system.


\vskip 0.2in \centerline{\bf \ \ Acknwoledgements} \vskip 0.2in
Our thanks go to the Isfahan University of Technology and
Institute for Studies in Theoretical Physics and Mathematics for
the financial support they made available to us.

\vskip 0.2in
\centerline{\bf \ \  References} \vskip 0.1in

\noindent [1] \ M. R. Douglas and N. A. Nekrasov, Rev. Mod. Phys.
{\bf 73}(2001)977-1029.

\noindent \ \ \ \ \  hep-th/0106048.

\noindent [2] \ R. Szabo, hep-th/0109162.

\noindent [3] \ B. DeWitt, in Gravitation, edited by L. Witten,
pp. 266-381(1962).

\noindent [4] \ H. Falomir et al, Phys. Rev. {\bf D66}\ (2002)\
045018, hep-th/0203260.

\noindent [5] \ M. Chaichian et al, Phys. Rev. Lett. {\bf 86} \
(2001)\ 2716, hep-th/0010175.

\noindent [6] \ S. Benczik et al, Phys. Rev. {\bf D66}\ (2002)\
026003, hep-th/0204049.

\noindent [7] \ L. N. Chang et al, Phys Rev. {\bf D65}\ (2002)\
125028, hep-th/0201017.

\noindent [8] \ S. Pireaux, J. P. Rozelot, and S. Godier,
astro-ph/0109032.

\noindent [9] \ Allen's Astrophysical Quantities, 4th
edition,edited by A.N. Cox

\noindent \ \ \ \ \ (Springer-Verlag, 2000)

\noindent [10] \ D.E. Groom et al.[Particle Data Group
collaboration ], Eur. Phys. J. C {\bf 15}, 1 \ (2000).

\vfill\eject

\bye